\documentclass[preprint2,tighten]{aastex}

\usepackage{graphicx}
\usepackage[below]{placeins} 

\singlespace

\bibpunct{(}{)}{;}{a}{}{,}

\shorttitle{Fossil group luminosity function}
\shortauthors{Mendes de Oliveira, Cypriano \& Sodr\'e Jr.}

\def\sol{\mbox{$_{\odot}$}} % solar-mass symbol
\def\Msol{\hbox{M$_{\odot}$}}

\def\h50{\hbox{h$_{50}$}}
\def\m12{\hbox{$\Delta$m$_{12}$}}

\newbox\grsign \setbox\grsign=\hbox{$>$} \newdimen\grdimen
\grdimen=\ht\grsign
\newbox\simlessbox \newbox\simgreatbox
\setbox\simgreatbox=\hbox{\raise.5ex\hbox{$>$}\llap
     {\lower.5ex\hbox{$\sim$}}}\ht1=\grdimen\dp1=0pt
\setbox\simlessbox=\hbox{\raise.5ex\hbox{$<$}\llap
     {\lower.5ex\hbox{$\sim$}}}\ht2=\grdimen\dp2=0pt
\newcommand\simgreat{\mathrel{\copy\simgreatbox}}

\newbox\simppropto
\setbox\simppropto=\hbox{\raise.5ex\hbox{$\sim$}\llap
     {\lower.5ex\hbox{$\propto$}}}\ht2=\grdimen\dp2=0pt

\begin{document}

\title{
The luminosity function of the fossil group RX~J1552.2+2013
\thanks{Based on observations obtained at the Gemini Observatory,
which is operated by the Association of Universities for Research in
Astronomy, Inc., under a cooperative agreement with the NSF on behalf of
the Gemini partnership: the National Science Foundation (United States),
the Particle Physics and Astronomy Research Council (United Kingdom),
the National Research Council (Canada), CONICYT (Chile), the Australian
Research Council (Australia), CNPq (Brazil) and CONICET (Argentina) --
Observing run ID: GN-2004B-Q-47.}
}

\author{Claudia L. Mendes de Oliveira}
\affil{Departamento de Astronomia, Instituto de Astronomia, Geof\'{\i}sica
e Ci\^encias Atmosf\'ericas da USP, Rua do Mat\~ao 1226, Cidade
Universit\'aria, 05508-090, S\~ao Paulo, Brazil}
\email{oliveira@astro.iag.usp.br}

\author{Eduardo S. Cypriano}
\affil{Southern Astrophysics Research Telescope, Casilla 603, La Serena, Chile
and Laborat\'orio Nacional de Astrof\'{\i}sica, 
CP 21, 37500-000 Itajub\'a - MG, Brazil}
\email{ecypriano@ctio.noao.edu}

\author{Laerte Sodr\'e Jr.}
\affil{Departamento de Astronomia, Instituto de Astronomia, Geof\'{\i}sica
e Ci\^encias Atmosf\'ericas da USP, Rua do Mat\~ao 1226, Cidade
Universit\'aria, 05508-090, S\~ao Paulo, Brazil}
\email{laerte@astro.iag.usp.br}

%\date{Received ? / Accepted ?}

\begin{abstract}
{
We determine the first fossil group luminosity function based on
spectroscopy of the member galaxies.  The fossil group RX~J1552.2+2013 has
36 confirmed members, it is at a mean redshift of 0.136 and has a velocity
dispersion of 623 km s$^{-1}$ (or 797 km s$^{-1}$ if four emission lines
galaxies in the outskirts of the velocity distribution are included).
The luminosity function of RX~J1552.2+2013, measured within
the inner region of the system ($\sim 1/3 R_{vir}$), in the range
--23.5 $<$ M$_{i^\prime}$ $<$ --17.5, is well fitted by a Schechter
function with M$_{i^\prime}^* = -21.3 \pm 0.4$ and $\alpha = -0.6 \pm
0.3$ or a Gaussian function centered on M$_{i^\prime} = -20.0\pm0.4$
and with $\sigma=1.29\pm 0.24$ i$^\prime$ mag (H$_0$ = 70  km s$^{-1}$
Mpc$^{-1}$, $\Omega_M$=0.3, $\Omega_{\Lambda}$=0.7).  The luminosity
function obtained from a photometric survey in g$^\prime$, r$^\prime$,
i$^\prime$-bands (and statistical background correction) confirms the
spectroscopically determined results.  There is a significant dip in the
luminosity function at M$_{r'}$=--18 mag, as also observed for other
clusters.  RX~J1552.2+2013 is a rich, strongly red-galaxy dominated
system, with at least 19 galaxies with magnitudes between m$_3$ and
m$_3+2$, within a surveyed circular area of radius 625 h$_{70}^{-1}$
kpc centered on the peak of the x-ray emission.  Its mass, $\sim 3
\times 10^{14}$ M$_\odot$, M/L of 507 M\sol/L$_B$\sol ~  and L$_X$ of
6.3 $\times$ 10$^{43}$ h$^{-2}_{50}$ ergs s$^{-1}$ (bolometric) are more
representative of a fossil cluster than of a fossil group.  The central
object of RX~J1552.2+2013 is a cD galaxy which may have accreted the more
luminous ($\sim$ L*) former members of the group. Although dynamical
friction and subsequent merging are probably the processes responsible
for the lack of bright galaxies in the system, for the fainter members,
there must be another mechanism  in action (perhaps tidal disruption) to
deplete the fossil group from intermediate-luminosity galaxies (M$_{r'}$
$\sim$ -18).  } 
\end{abstract}

%6 keywords at maximum!  %

\section{Introduction}

 A fossil group is a system with a large extended X-ray halo, dominated by one
brighter-than-L* elliptical galaxy which is surrounded by low-luminosity
companions (where the difference in magnitude between the bright dominant
elliptical and the next brightest companion is $>$ 2 mag).  The first system
identified as a fossil group was RXJ1340.6+4018, at a redshift of 0.171 (Ponman
et al.\ 1994).  A few years later, \citet{vikhlinin99} catalogued four systems
which they called ``X-ray overluminous elliptical galaxies", given the
unusually extended X-ray halos observed for what seemed to be isolated objects.
These had essentially the same properties as fossil groups \citep[in fact one
of the four systems was the prototype fossil group from][]{ponman94}.  
Subsequently, \citet{jones03} catalogued five other fossil groups, selected
from the WARPS survey. More recently, \citet{yoshioka04} catalogued four nearby
systems  (of which one is not a fossil group, namely RX~J0419.6+0225 is the
rich group NGC 1550). A few other individual fossil groups were recently
catalogued: \citet{khosroshahi04}, \citet{sun04} and \citet{ulmer05} presented
detailed X-ray studies of three additional fossil groups, NGC~6482, ESO~3060170
and Cl 1205+44, respectively. In total, there are then 15 known fossil groups
catalogued to date (see Table \ref{allfossil}). The X-ray halo of a fossil
group is comparable in luminosity and extension to that of a rich group or poor
cluster, suggesting that their masses may also be similar. 
Such systems are also thought 
to be quite abundant (their density is $\sim$ 2.4
$\times$ 10$^{-7}$ Mpc$^{- 3}$, Jones et al. 2003) 
and therefore they may yield an important
contribution to the luminosity density and to the baryon budget of the
universe.

The main scenario for the formation of fossil groups is that they are possibly
the successors of massive versions of today's compact groups \citep[\it merging
origin, \normalfont][]{jones03}.  In such a scenario, the brightest galaxies of
the system would have merged, leaving behind a group/cluster dominated by a
single elliptical galaxy, surrounded by small companions.
On the other hand, a recent X-ray study by \citet{yoshioka04} points
out that the M/L of four fossil groups studied by them reach up to
M$_{200}$/L$_B$ = 1100 M$_\odot$/L$_\odot$, being at least one order of
magnitude higher than the typical M/L for groups/clusters of galaxies of
similar mass, which makes it very unlikely that fossil groups could be
descendants of groups/clusters.  However, other recent works derive much lower
values of M/L for other such systems  (e.g. \citeauthor{sun04} 2004) 
and therefore
this matter is not yet settled.  In addition, it is possible that there is more
than one formation mechanism for fossil groups.  In the formation scenario
proposed by \citet{yoshioka04}, fossil groups are the massive end of the
elliptical galaxy distribution, formed at high redshift ({\it fossil-elliptical
origin}).

In a recent paper, \citet{donghia04}, based on the luminosity function of one
such group, pointed out that fossil groups may pose a severe problem for the
cold dark matter models. They pointed out that these systems do not have as
much substructure as expected for such a massive system, and moreover the
``missing galaxies" are much more luminous than those lacking in poor groups
and in the field.  However, the luminosity function of one fossil group in
which that conclusion is based has been determined through imaging of the group
and statistical background subtraction \citep{jones00}, with very few known
redshifts. In fact, most of the past studies about fossil groups  focused on
their X-ray properties and very little information on their optical properties
is available. In this paper, we present the first secure luminosity function
determination of a fossil group, using spectroscopic data obtained with
GMOS \citep{GMOS} on the Frederick C. Gillett Telescope (Gemini North) at Mauna
Kea.

We present a spectroscopic study of RX J1552.2 +2013, 
one of the groups originally
catalogued by \citet{jones03}, selected from an X-ray survey done with ROSAT
(WARPS) with well defined selection criteria. This system has high 
bolometric X-ray-luminosity
\citep[6.3 x 10$^{43}$ h$_{50}^{-2}$ erg\,s$^{-1}$, ][]{jones03} 
and, before this study,
it had one of the largest number of known members (four). 
We determine the
luminosity function of the group from g$^\prime$, r$^\prime$, i$^\prime$
photometry down to i$^\prime$=23 and spectroscopy of 36 members. Sections 2 and
3 describe 
the observations, the reduction procedure and the results. In Section 3
we include, in Table \ref{allfossil}, a summary of the main properties of the
15 known fossil groups to date. In section 4 we discuss the implications of our
results for the understanding of the nature and evolution of fossil groups.

When needed, we adopt the following values for the cosmological parameters:
H$_0$ = 70  km s$^{-1}$ Mpc$^{-1}$, $\Omega_M$=0.3, and 
$\Omega_{\Lambda}$=0.7.

%-------------------------------------------------------------------

\section{\label{obs}Observations and reductions}

The imaging and multi-slit spectroscopic observations of the group RX~J1552.2+2013 were
done with the GMOS instrument, mounted on the Gemini North telescope, on Aug
8th and Sep 9th/2004 respectively.

\begin{figure}[h!]
\includegraphics[width=1.0\columnwidth]{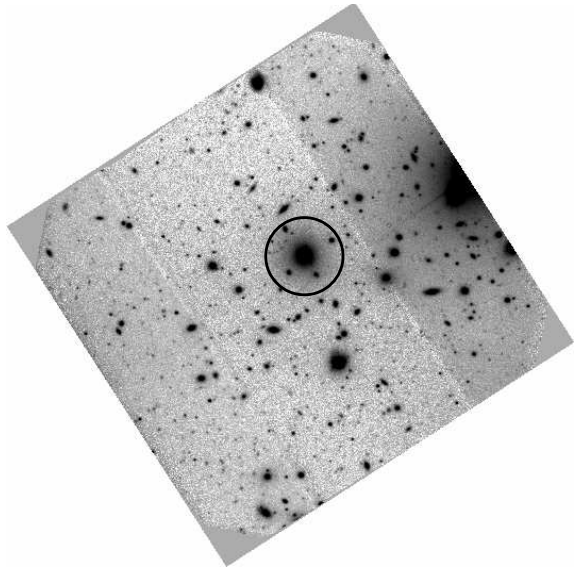}
\caption{\label{image}Optical  image of RX~J1552.2+2013. The field of view is 5.6 arcmin 
on a side, or 803 h$_{70}^{-1}$ kpc  at the object redshift. North is
up and east is to the left. The circle indicates the isophote where the
surface brightness of the galaxy is equal to one sigma of the sky}
\end{figure}

The imaging consisted of $3\times200$s exposures in each of the three filters
from the SDSS system \citep{sloan} g$^\prime$, r$^\prime$ and i$^\prime$. The
typical FWHM for point sources was $\sim$ 0.75" in all images. All observations
were performed in photometric conditions.  Fig. \ref{image} displays the
r$^\prime$ image of the system.

The calibration to the standard SDSS system  was made with the
general extinction coefficients provided by the Gemini 
observatory~\footnote{\texttt{www.gemini.edu/sciops/instruments/gmos/gmosPhotStandards.html}}.
The accuracy of the calibration is claimed to be  within 5\% to 8\%.

Standard reduction using the Gemini package GMOS was used. After flat fielding
and cleaning the images for cosmic rays,  the final frames were analyzed with
the program SExtractor \citep{ber96}. Positions and magnitudes (isophotal and
aperture) were obtained for all objects. We estimate that the galaxy catalog is
essentially complete down to  23.5 i$^\prime$ magnitude,
since the number counts turn over at i'=24 mag.

\begin{figure}[h!]
\includegraphics[width=1.0\columnwidth]{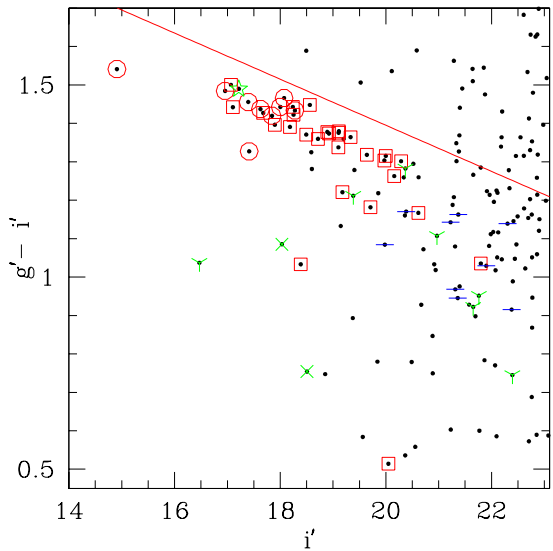}
\caption{\label{cmd}
Color-magnitude diagram of the galaxies in the RX~J1552.2+2013 field. Points marked with
squares (members), `Y'
% `{\large $\Ydown$}'
 (non-members) and  `--' (with spectra
but no redshift) represent the galaxies observed spectroscopically  by us with
GMOS-N. The circles (members), `$\times$' (non-members) and
stars (doubtful cases) represent galaxies
observed by Jones (2004, priv. comm.). The line indicates an upper limit 
for the cluster red-sequence we adopted when selecting the spectroscopic
targets.}
\end{figure}

Candidates for spectroscopy were chosen based on the color-magnitude
diagram shown in Fig. \ref{cmd}: most of the galaxies with M$_i$ $<$
-20, below the red sequence (see continuous line in the figure) were
observed spectroscopically.  Galaxies above  this line are expected to
be in the background, since their colors are redder than the expected
colors of elliptical galaxies at the group redshift.  Note that the
outermost observed galaxy which turned out to be a member of the
group/cluster has a distance of 625 h$_{70}^{-1}$ kpc from the X-ray
center of RX~J1552.2+2013.

\begin{figure*}[tb!]
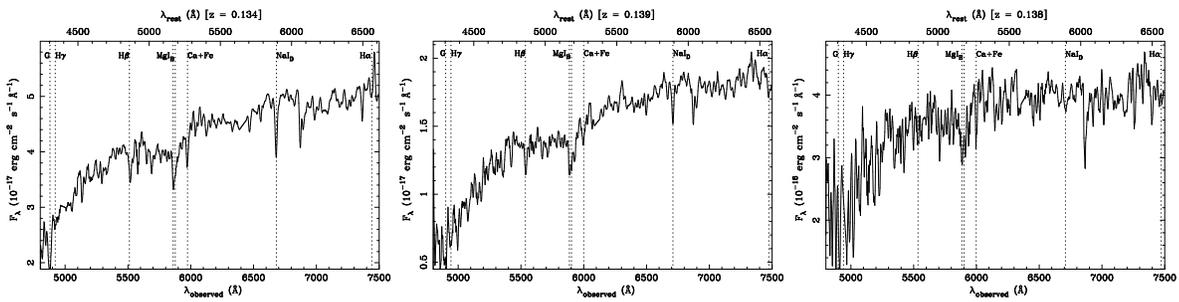

\includegraphics[width=0.67\columnwidth]{f3a.eps} %
%\vskip 0.3cm
\includegraphics[width=0.67\columnwidth]{f3b.eps} %
%\vskip 0.3cm
\includegraphics[width=0.67\columnwidth]{f3c.eps} 
\caption{\label{spectra}
Spectra of member galaxies. The panels show, from left to right, spectra
of galaxies G06.8+1312, G07.6+1439 and G11.4+1509 as examples of the
data taken for this study.  These spectra are the ones for which the
largest, median and lowest cross-correlation  coefficient was measured,
respectively.  The wavelength range used for the the cross-correlation
(4800--7500\AA) is shown. In this figure, 
the spectra have been smoothed using a boxcar
filter of size 13.6 \AA~(5 pixels), for the sake of clarity.
}
\end{figure*}

One single multi-slit exposure of 100 minutes was obtained through a mask with
1.0\arcsec ~ slits, using the R400 grating, for a final resolution of 8 \AA~(as
measured from the FWHM of the arc lines), covering aproximatelly the range 4000
-- 8000 \AA~ (depending on the position of each slitlet). The spectra of three
selected galaxies (from the best to the worst signal to noise) are shown in
Fig. \ref{spectra}.  Only four emission-line galaxies turned out to be members
of the fossil-group. Most of them lie in the outskirts of the galaxy velocity
distribution (see Table \ref{tab_spectra}).

 Standard procedures were used to reduce the multi-slit spectra using tasks
within the Gemini {\sc IRAF} \footnote{IRAF is distributed by the National
Optical Astronomy Observatories, which are operated by the Association of
Universities for Research in Astronomy, Inc., under cooperative agreement with
the National Science Foundation.} package. Wavelength calibration was done
using Cu-Ar comparison-lamp exposures before and after the exposure. 

Redshifts for galaxies with absorption lines were determined using the
cross-correlation technique \citep{T&D} as implemented in the package {\sc
RVSAO} \citep{rvsao} running under {\sc IRAF}. The final heliocentric
velocities of the galaxies were obtained by cross-correlation with several
template spectra. The final errors on the velocities were determined from the
dispersion in the velocity estimates using several different galaxy and star
templates. In the case of the three  emission-line redshifts, the error was
estimated
from the dispersion in redshifts obtained using different emission lines.  The
resulting heliocentric velocities typically have estimated rms errors between
25 and 100 km s$^{-1}$.  The S/N of the data,  measured at the continuum region around
6000--6300\AA, ranged from 10 to 30.

Table \ref{tab_spectra} lists positions, isophotal magnitudes, 
aperture (g' - i') colors, radial velocities and the Tonry \& Davis
cross-correlation coefficient R
for all galaxies with reliable velocity determination obtained in
this study.

\begin{deluxetable}{lccccccc}
\tablewidth{0pt}
\tablecaption{Spectral data for galaxies in the RX~J1552.2+2013 field. \label{tab_spectra}}									   
\tablehead{\colhead{(1)} & \colhead{(2)} & \colhead{(3)} & \colhead{(4)} &
\colhead{(5)} & \colhead{(6)} & \colhead{(7)}\\
\colhead{Name} & \colhead{RA (2000)} & \colhead{DEC (2000)} &
\colhead{i$^\prime$ (AB Mag.)} &
\colhead{g$^\prime$-i$^\prime$} & \colhead{Vel. (km s$^{-1}$)} & \colhead{~R}}													   
\startdata						   
G04.5+1221\tablenotemark{a} & 15 52 04.5 & +20 12 21 & 19.38 & 1.21 & $ ~26819 \pm ~32$ & \nodata\tablenotemark{b}\\
G16.4+1605 & 15 52 16.4 & +20 16 05 & 16.47 & 1.04 & $ ~26820 \pm ~18$ & \nodata\tablenotemark{b}\\
\\
G12.2+1439 & 15 52 12.2 & +20 14 39 & 18.49 & 1.37 & $ ~38715 \pm ~25$ &  6.3			 \\
G23.2+1452 & 15 52 23.2 & +20 14 52 & 20.05 & 0.51 & $ ~39290 \pm ~36$ & \nodata\tablenotemark{b}\\
G10.0+1342 & 15 52 10.0 & +20 13 42 & 20.61 & 1.17 & $ ~39374 \pm 100$ & \nodata\tablenotemark{b}\\
G12.2+1534 & 15 52 12.2 & +20 15 34 & 18.18 & 1.39 & $ ~39538 \pm ~43$ &  7.3			 \\
G06.4+1406 & 15 52 06.4 & +20 14 06 & 19.71 & 1.18 & $ ~39765 \pm 100$ &  3.2			 \\
G06.8+1312 & 15 52 06.8 & +20 13 12 & 17.07 & 1.50 & $ ~40103 \pm ~34$ & 10.2			 \\
G09.4+1157 & 15 52 09.4 & +20 11 57 & 18.72 & 1.36 & $ ~40115 \pm ~46$ &  6.4			 \\
G10.9+1224 & 15 52 10.9 & +20 12 24 & 21.80 & 1.03 & $ ~40206 \pm ~59$ &  2.8			 \\
G12.9+1305 & 15 52 12.9 & +20 13 05 & 19.33 & 1.36 & $ ~40211 \pm ~48$ &  5.2			 \\
G16.0+1009 & 15 52 16.0 & +20 10 09 & 17.10 & 1.44 & $ ~40227 \pm ~45$ &  5.9			 \\
G15.9+1211 & 15 52 15.9 & +20 12 11 & 19.10 & 1.34 & $ ~40305 \pm ~44$ &  5.6			 \\
G11.9+1234 & 15 52 11.9 & +20 12 34 & 19.10 & 1.38 & $ ~40344 \pm ~39$ &  5.8			 \\
G08.7+1136 & 15 52 08.7 & +20 11 36 & 18.24 & 1.44 & $ ~40472 \pm ~56$ &  6.7			 \\
G05.9+1246 & 15 52 05.9 & +20 12 46 & 19.18 & 1.22 & $ ~40704 \pm ~47$ &  3.2~\tablenotemark{c}  \\
G15.9+1039 & 15 52 15.9 & +20 10 39 & 17.68 & 1.43 & $ ~40748 \pm ~39$ &  6.5			 \\
G17.0+1406 & 15 52 17.0 & +20 14 06 & 20.00 & 1.32 & $ ~40853 \pm ~66$ &  4.1			 \\
G18.6+1410 & 15 52 18.6 & +20 14 10 & 19.64 & 1.32 & $ ~41001 \pm ~98$ &  4.1			 \\
G16.8+1208 & 15 52 16.8 & +20 12 08 & 19.97 & 1.30 & $ ~41126 \pm ~87$ &  3.3			 \\
G14.9+1404 & 15 52 14.9 & +20 14 04 & 18.25 & 1.42 & $ ~41142 \pm ~24$ &  7.4			 \\
G10.5+1610 & 15 52 10.5 & +20 16 10 & 17.90 & 1.40 & $ ~41187 \pm ~35$ &  7.2			 \\
G14.4+1115 & 15 52 14.4 & +20 11 15 & 20.16 & 1.26 & $ ~41275 \pm ~40$ &  3.5			 \\
G11.4+1509 & 15 52 11.4 & +20 15 09 & 20.29 & 1.30 & $ ~41503 \pm ~96$ &  2.4			 \\
G07.6+1439 & 15 52 07.6 & +20 14 39 & 18.92 & 1.37 & $ ~41582 \pm ~45$ &  5.9			 \\
G16.9+1531 & 15 52 16.9 & +20 15 31 & 18.89 & 1.38 & $ ~41609 \pm ~48$ &  6.6			 \\
G10.7+1519 & 15 52 10.7 & +20 15 19 & 19.11 & 1.38 & $ ~41797 \pm ~97$ &  3.8			 \\
G13.2+1326 & 15 52 13.2 & +20 13 26 & 18.56 & 1.45 & $ ~42041 \pm ~48$ &  7.2			 \\
G07.8+1503 & 15 52 07.8 & +20 15 03 & 18.39 & 1.03 & $ ~42045 \pm ~74$ & \nodata\tablenotemark{b}\\\\
G18.8+1345 & 15 52 18.8 & +20 13 45 & 20.97 & 1.11 & $ ~49262 \pm ~54$ &  2.0			 \\
G11.4+1359 & 15 52 11.4 & +20 13 59 & 20.37 & 1.28 & $ ~68668 \pm ~40$ &  4.5			 \\
G10.6+1139 & 15 52 10.6 & +20 11 39 & 21.76 & 0.95 & $ ~93209 \pm ~64$ & \nodata\tablenotemark{b}\\
G12.8+1516 & 15 52 12.8 & +20 15 16 & 21.65 & 0.92 & $ 101640 \pm ~56$ & \nodata\tablenotemark{b}\\
G07.4+1239 & 15 52 07.4 & +20 12 39 & 22.40 & 0.75 & $ 105743 \pm 101$ & \nodata\tablenotemark{b}\\
\enddata
\tablenotetext{a}{The names of the galaxies are based on their 2000 celestial coordinates
(RA seconds and DEC minutes and seconds). Thus galaxy Gab.c+defg is located at
15 52 ab.c +20 de fg.}
\tablenotetext{b}{Redshift measured from emission lines.}
\tablenotetext{c}{Redshift measured from absorption lines but it also has 
emission lines in the spectrum.}	
\end{deluxetable}

\section{\label{res}Results}

\subsection {Galaxy velocity distribution}

Using the heliocentric radial velocities listed in Table
\ref{tab_spectra}, we define the putative members of RX~J1552.2+2013
as the 27 galaxies with velocities between 38715 and 42045 km s$^{-1}$.
There are nine other confirmed members studied by Jones et al. (private
communication) which were included in most of our analysis, but not in the
galaxy velocity distribution and determination of dynamical mass.

\begin{figure}[h!]
\includegraphics[width=1.0\columnwidth]{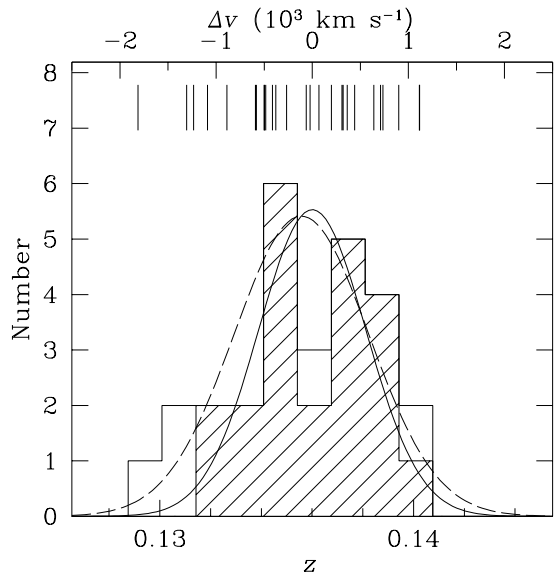}
\caption{\label{veldisp}
Velocity histogram of RX~J1552.2+2013.  It shows the distribution of the radial velocities
of 27 galaxies in the inner 625 h$_{70}^{-1}$ kpc  radius of RX~J1552.2+2013, with
redshifts within $\pm$2500 km/s of the systemic velocity of the group. The
sticks on the upper part of the plot show velocities of individual objects.  
The {\sc rosat} bi-weighted estimator gives a velocity dispersion  $\sigma =
797$ km/s and a redshift of $\langle z \rangle = 0.1357$ (dashed line), for the
sample of 27 objects. The dashed part shows a sub-sample of 22 galaxies,
obtained by  excluding four galaxies with emission lines and one other galaxy
that becomes a $\pm$3 $\sigma$-outlier with this exclusion.  The bi-weighted
estimator returned $\sigma=635$ km/s and $\langle z \rangle = 0.136$
(continuous line) for this smaller subsample.}
\end{figure}

The data were analyzed with the statistical
software {\sc rostat} \citep{beers}, which did not find any large gap
in the velocity distribution. In addition, no other data-points were found
outside a $\pm3\sigma$ range.  Fig. \ref{veldisp} shows the velocity
histogram for the 27 member galaxies studied by us.

Using the robust bi-weighted estimator, {\sc rostat}, the following values for
the systemic redshift and velocity dispersion were found: $\langle z \rangle =
0.1357\pm0.0011$ and $\sigma = 797\pm185$ km s$^{-1}$,  respectively. 

It is interesting to note, however, the effect of the exclusion of the
emission-line galaxies when deriving these quantities, since those objects are
believed to be recently accreted by the cluster, thus not yet virialized
\citep[e.g.][]{sodre89,biviano97}.  Indeed, a simple inspection of Table
\ref{tab_spectra} shows that 3 out of 4 emission-line galaxies occupy the 
edges of the velocity distribution.

The exclusion of the emission-line galaxies reduces the velocity dispersion of
the group. As a consequence, the radial velocity for one  other galaxy,
G12.2+1439 falls outside the new $\pm3\sigma$ range. If this galaxy is also
excluded, the new subsample is then composed of 22 galaxies, for which the
bi-weighted values are $\langle z \rangle = 0.136\pm0.001$ and $\sigma =
635\pm164$ km s$^{-1}$.  We consider this determination of redshift and velocity
dispersion the most reliable ones for RX~J1552.2+2013  (see below). 
Note that for the determination of the luminosity function all 36 galaxies in
the sample were used (including the ones with emission lines).

\subsubsection{Dynamical mass}

We determine the dynamical mass of the cluster by using four different mass
estimators, as suggested by \citet{heisler85}: virial, projected, average and
median mass estimators (see Table \ref{mass}).

The maximum difference between different mass estimates is about 24\%.
According to \citeauthor{heisler85} there is a 75\% chance that the derived
mass of any single system is within a factor of order 2 of the correct value. 
However, caution notes have been given by several authors
\citep[e.g.][]{girardi} on the reliability of mass determinations when the
galaxies are distributed over a small (central) part of the cluster and the
number of redshifts is limited. The final, adopted mass for the cluster was
obtained from the average value of the results of the four different
estimators.

\begin{deluxetable}{lcccc}
\tablewidth{0pt}
\tablecaption{Mass Estimates \label{mass}}									   
\tablehead{\colhead{(1)} & \colhead{(2)} & \colhead{(3)} & \colhead{(4)} &
\colhead{(5)}\\
\colhead{} & \multicolumn{2}{c}{Non emission-line galaxies} &
\multicolumn{2}{c}{All galaxies} \\											   
\colhead{Estimator} & \colhead{Mass ($10^{14}$ \Msol)} &
\colhead{M/L$_B$ (\Msol/L$_B$\sol)} & \colhead{Mass ($10^{14}$ \Msol)} &
\colhead{M/L$_B$ (\Msol/L$_B$\sol)}\\}
\startdata
Virial     & 2.64 & 521 & 4.11 & 811 \\
Projected  & 2.73 & 539 & 4.10 & 809 \\
Average    & 2.20 & 433 & 3.29 & 649 \\
Median     & 2.71 & 534 & 3.74 & 738 \\
           &      &     &      &     \\
Mean value & 2.57 & 507 & 3.81 & 752 \\	      
\enddata
\end{deluxetable}
 
%\FloatBarrier

\subsubsection {Location in the L$_X$--$\sigma$ relation}

The measured value of the bolometric X-ray luminosity of RX~J1552.2+2013 is
$6.31\times10^{43}$ h$_{50}^{-2}$ erg\,s$^{-1}$ \citep{jones03} and there is no 
available measurement of T$_X$ in the literature.
This value of L$_X$ is slightly above the value found for groups, L$_X \la 4
\times 10^{43}$ h$_{50}^{-2}$ erg\,s$^{-1}$
and $\sigma \la 600$ km s$^{-1}$ \citep{xue00}, being more consistent
with a low-mass cluster.

According to the relations for groups and clusters from \citet{mahdavi01}, we
find that for a system with this X-ray luminosity we would expect a
velocity dispersion of about $\sigma$ = 534 km s$^{-1}$, in good agreement with our
direct measurement. In fact, even the value for the velocity dispersion
calculated including emission-line galaxies (797 km s$^{-1}$) is within the
dispersion of the points in their Fig. 1--(b).

The characterization of RX~J1552.2+2013 as a cluster is also supported by the richness of
this system. Within 625 h$_{70}^{-1}$ kpc of the central galaxy there are 19
galaxies between m$_3$ (magnitude of the third brightest member) and m$_3+2$.
The \citet{aco} definition for a cluster is at least 30 galaxies within a
radius of 1.5 Mpc. Given the high density of observed members, this selection
criterium would certainly have been fulfilled, had we surveyed a larger area.

\subsubsection{Mass-to-light ratio} 
  
We estimated the luminosity of the central regions of the group by adding
up the luminosities of the 36 spectroscopically confirmed members,  taking
into account the completeness correction derived from  the spectroscopic
sampling (see Fig. \ref{flum}).The completeness correction was
defined in the following way: $N(m) = N_{grp}(m) / C(m)$, where $N_{grp}$
is the number of galaxies spectroscopically confirmed as members in a
given magnitude bin. $C(m)$ is the completeness, defined as:
$N_{vel}(m)/N_{tot}(m)$, the number of galaxies in a given magnitude bin
for which we were able to get a reliable redshift, over the total number
of galaxies in the same magnitude bin. We note that, for this analysis,
we have negleted all the galaxies redder than the red-cluster sequence,
shown in Fig. \ref{cmd}.

In order to compare with previous results, we have transformed our SLOAN
g$^\prime$ and r$^\prime$ magnitudes in standard Johnson--Morgan B  
magnitude using the transformations given in \citet{sloan}. The total
magnitudes have been corrected for Galactic--extinction
\citep{schlegel} as well as for the $k$-correction   
\citep{fukugita95}, under the assumption that all
galaxies are early--type, which is valid for most of them.

The total luminosity calculated within 625 kpc h$_{70}^{-1}$ is $5.06\times
10^{11}$ \Msol$_B$. This leads to a mass-to-light ratio in units of
M\sol/L$_B$\sol ~ of 507.  Results for the several mass estimates 
and for the samples with and without emission-line galaxies are
presented in Table \ref{mass}.

\subsection{The Luminosity Function}

\begin{figure*}[tb!]
\includegraphics[width=0.67\columnwidth]{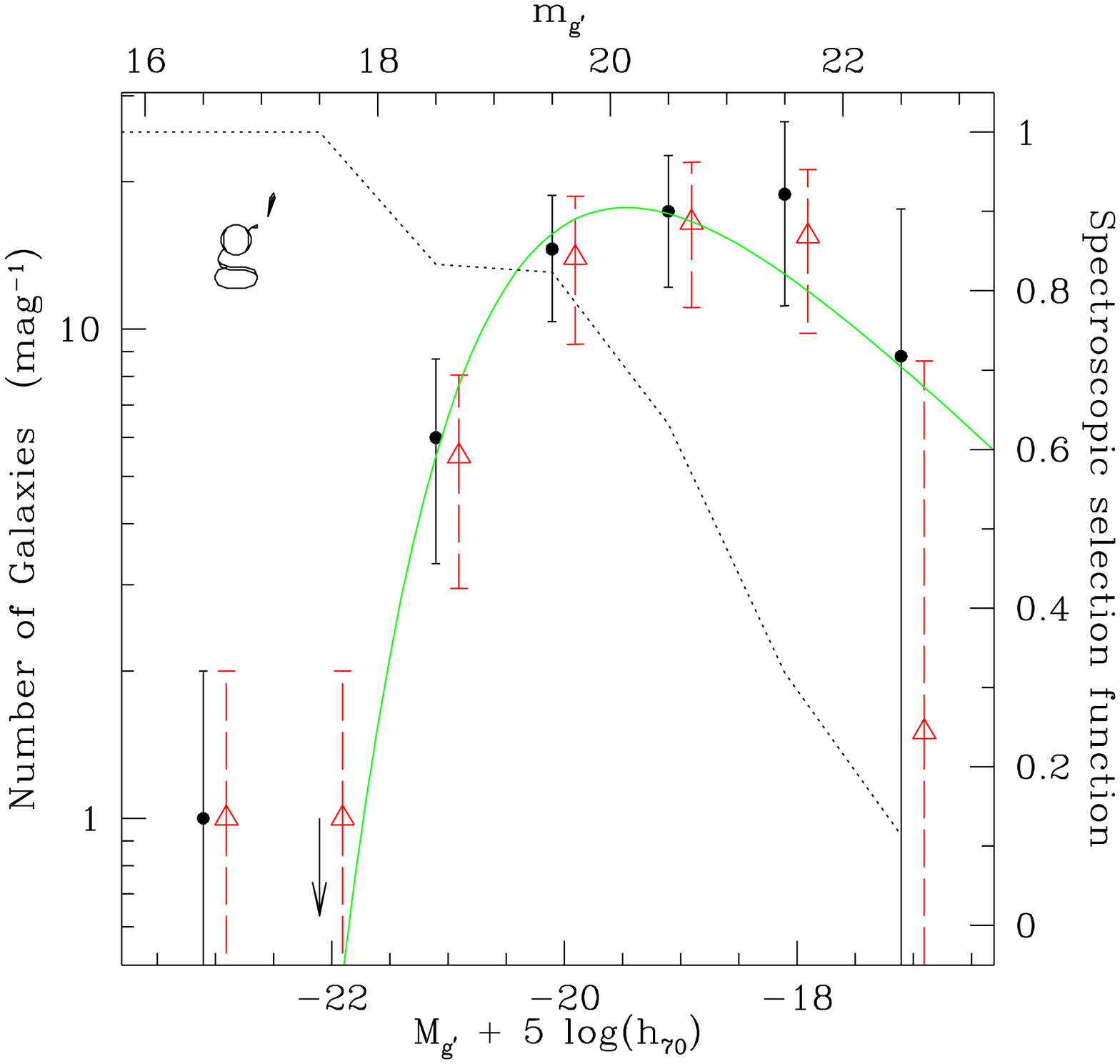}
\includegraphics[width=0.67\columnwidth]{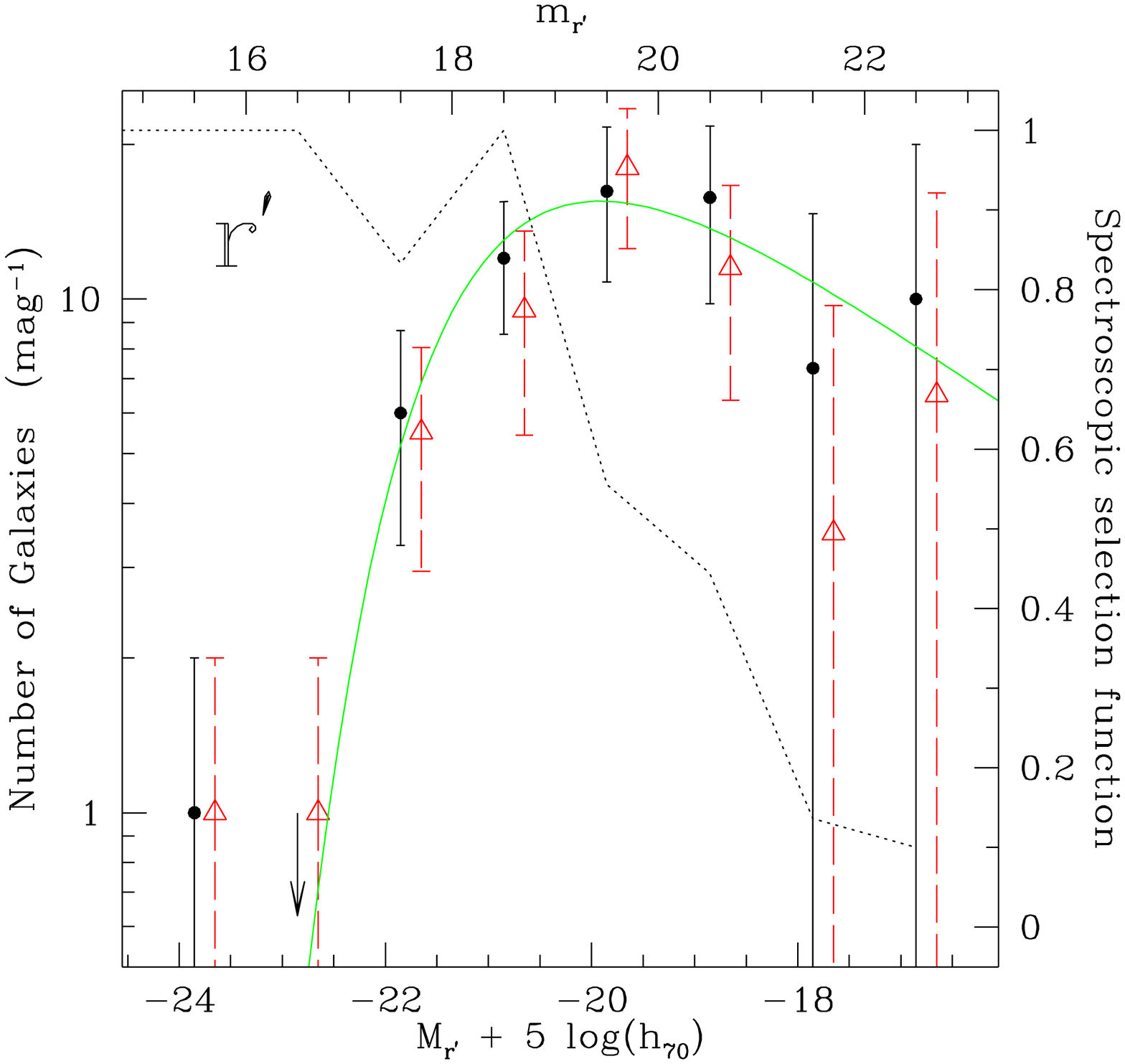}
\includegraphics[width=0.67\columnwidth]{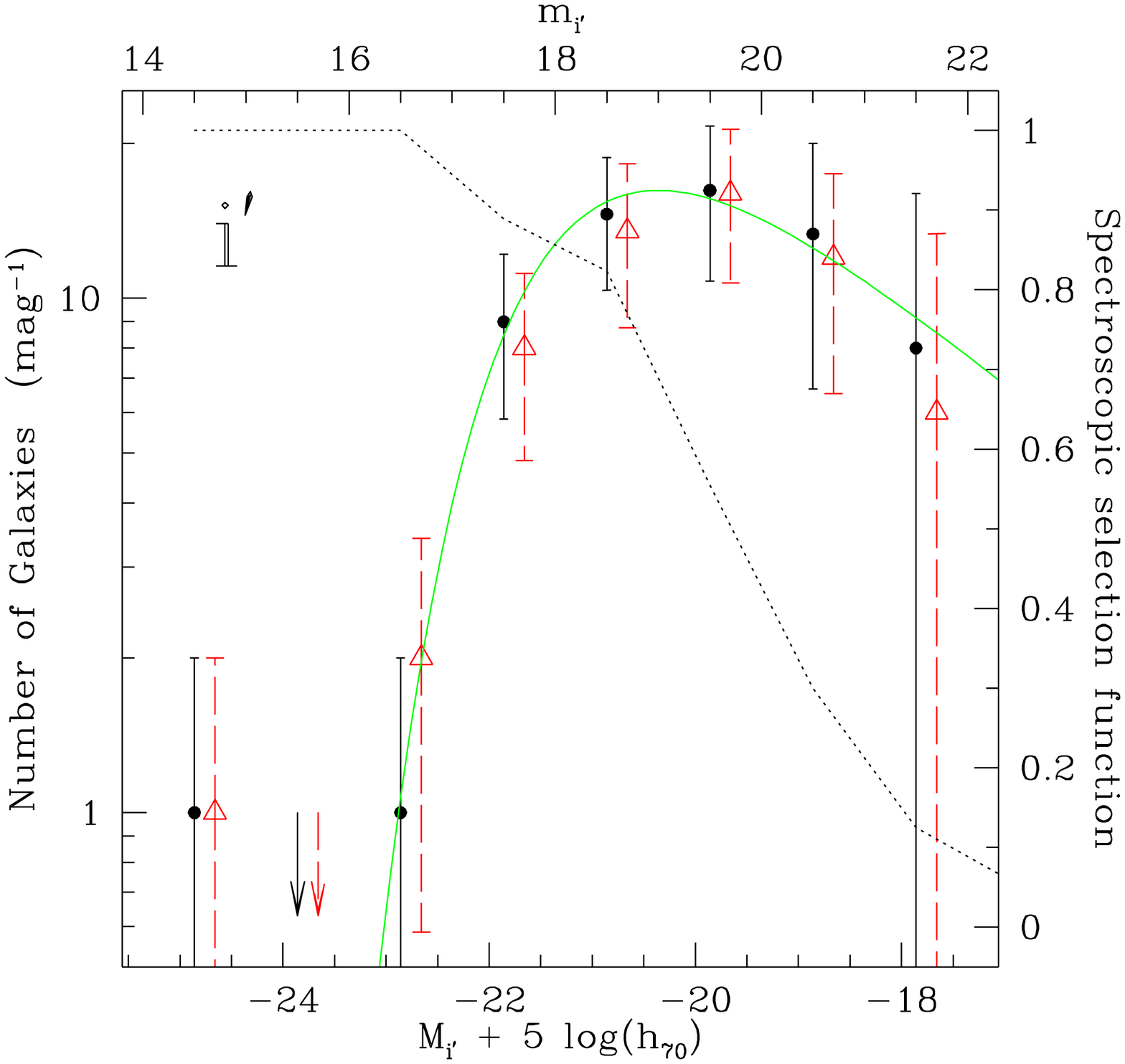}
\caption{\label{flum}
Luminosity Function of RX~J1552.2+2013.  The panels show, from left to right, the
luminosity functions in the g$^\prime$, r$^\prime$ and i$^\prime$ bands,
respectively. The solid circles show the completeness-corrected number of
spectroscopically confirmed members of  RX~J1552 per 1.0 magnitude bin in the
GMOS field. The error bars are 1$\sigma$ Poissonian errors. The arrows show
bins with number of galaxies less or equal to zero. The dotted line is the
selection function of the spectroscopic sample. The continuous lines show the
best fitted Schechter function of the spectroscopic sample (see values in Table
\ref{tab_FdL}). The brightest galaxy of the cluster was not included in
the fit. The open triangles show the photometrically-determined 
luminosity function estimated through number counts and statistical subtraction
of the background. The points have been shifted by 0.2 mag, for the sake of
clarity. The agreement between the spectroscopically and
photometrically-determined luminosity functions is excellent.}
\end{figure*}

We show in Fig. \ref{flum} the luminosity function of RX~J1552.2+2013 (solid circles) for
galaxies with spectroscopically confirmed membership either obtained in this
paper  (27 galaxies) or given by Jones et al.  (private communication, 9
galaxies), corrected for incompleteness.  The absolute magnitudes were
calculated after correcting the observed magnitudes for Galactic extinction and
applying $k$-corrections.
The selection function, also shown in Fig. \ref{flum}, was calculated
considering  only galaxies bluer than the upper limit of the adopted red 
cluster sequence. 

\begin{deluxetable}{ccccccc}
\tabletypesize{\scriptsize}
\tablecaption{Luminosity Function \label{tab_FdL}}									   
\tablehead{\colhead{(1)} & \colhead{(2)} & \colhead{(3)} & \colhead{(4)} &
\colhead{(5)} & \colhead{(6)} & \colhead{(7)}\\
 & & \multicolumn{2}{c}{Schechter Function} &
 \multicolumn{2}{c}{Gaussian Function} & \\       				    
 & \colhead{Band} & \colhead{M$^*+5\log(h_{70})$} &
\colhead{$\alpha$} & \colhead{$\langle M\rangle+5\log(h_{70})$} &
\colhead{$\sigma_M$} & \colhead{Magnitude range}}
\startdata
              & g$^\prime$ & $-20.05\pm0.45$  & $-0.42\pm0.32$ & $-19.02\pm0.32$ & $1.20\pm0.19$ &18.0 -- 23.0\\
Spectroscopic & r$^\prime$ & $-21.07\pm0.48$  & $-0.65\pm0.26$ & $-19.59\pm0.41$ & $1.31\pm0.25$ &17.0 -- 23.0\\
              & i$^\prime$ & $-21.34\pm0.40$  & $-0.59\pm0.29$ & $-19.99\pm0.42$ & $1.29\pm0.24$ &16.0 -- 22.0\\
\hline
               & g$^\prime$ & $-20.35\pm0.58$  & $-0.74\pm0.42$ &
	             $-19.01\pm0.50$ & $1.21\pm0.27$ & 18.0 -- 22.0\\
Spectroscopic$~\tablenotemark{a}$ & r$^\prime$ & $-21.18\pm0.57$  & $-0.77\pm0.37$ &
	                     $-19.71\pm0.44$ & $1.24\pm0.27$ & 17.0 -- 21.0\\
               & i$^\prime$ & $-21.38\pm0.44$  & $-0.63\pm0.35$ &
	                   $-20.10\pm0.42$ & $1.23\pm0.24$ & 16.0 -- 21.0\\
\hline
              & g$^\prime$ & $-20.21\pm0.70$  & $-0.47\pm0.42$ & $-19.14\pm0.32$ & $1.25\pm0.22$ &17.0 -- 23.0\\
Photometric   & r$^\prime$ & $-21.27\pm0.62$  & $-0.64\pm0.30$ & $-19.82\pm0.36$ & $1.31\pm0.27$ &16.0 -- 23.0\\
              & i$^\prime$ & $-21.59\pm0.51$  & $-0.67\pm0.27$ & $-20.03\pm0.40$ & $1.39\pm0.27$ &15.0 -- 23.0\\
\enddata
\tablenotetext{a}{Note that these results have more limited magnitude range since bins with only one galaxy, at the faint end,  were excluded}
\end{deluxetable}

We have also estimated photometrically the luminosity function of  RX~J1552.2+2013 down
to $\sim$ 23 mag, in the three bands 
(the photometric sample is 
complete  at this magnitude) by adopting the following procedure. First,
we consider only galaxies bluer than the  limit of the red sequence
in the color-magnitude diagram shown in Fig. \ref{cmd}.  We then binned
the magnitudes of  these galaxies in 1.0 magnitude bins and subtracted from
each bin the average number of field galaxies with colors also below the
red-sequence line, taken from two control fields. The two
high-galactic latitude empty fields
used as control fields were obtained with the same
telescope, instrument and filters (Boris et al. 2005, in preparation). The
counts derived from the two fields are in good agreement among themselves
and with those of the Hawaii HDF-N  
field \citep{capak04}. The errors in each bin
of the luminosity function were computed assuming Poissonian statistics
for  both field and group counts. These photometric luminosity functions
are also shown in Fig. \ref{flum}  as open triangles. They are strikingly
similar to the spectroscopic results (solid circles).
Note, however, that the determination
of the luminosity function using galaxy counts on and off the field is
significantly more uncertain than the spectroscopic determination given
cosmic variance.

\begin{figure}[!ht]
\includegraphics[width=1.0\columnwidth]{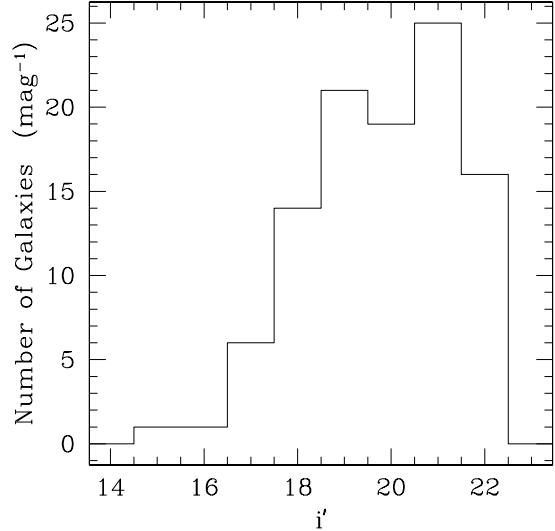}
\caption{\label{counts}
Number counts, per magnitude bin,
below the red cluster sequence of RX~J1552.2+2013 (i.e.
below the continuous line in Fig. \ref{cmd}).  
Note that the number counts level off at the
faint end indicating a flat or decreasing
luminosity function for this system.}
\end{figure}

The shape of the luminosity function is very similar in the three
bands. However, it is somewhat unexpected: the number of galaxies in the
group flattens and then decreases for magnitudes fainter than g$^\prime
\sim 20$, r$^\prime \sim 19.5$ and i$^\prime \sim 19$. 
This can be clearly seen in the histogram shown in Fig. \ref{counts},
of the number of galaxies, per magnitude interval,
below the red-cluster sequence.

The shapes of the luminosity functions shown in Fig. 5 are well
described by Gaussian functions.  In the i$^\prime$-band this Gaussian
is centered at M$_{i^\prime}= -20.0\pm0.4$ (m$_{i^\prime}=19.37$) and has
$\sigma=1.29\pm 0.3$  mag. If we attempt a fit with a Schechter function,
the best parameters are, in the i$^\prime$-band, M$_{i^\prime}^* =
-21.3 \pm 0.5$ (m$_{i^\prime}=18.05$) and $\alpha = -0.59 \pm 0.24$,
in the range 16 $<$ m$_{i^\prime}$ $<$ 22.  Results for all bands
and parameters for both Gaussian and Schechter luminosity functions are
presented in Table 3.  Note that the results for the spectroscopic
luminosity function do not change significantly if we restrict the magnitude
interval to exclude the faint magnitude bins which have one
galaxy only (see results in middle pannel of Table 3).  Lines indicating
the best Schechter luminosity functions fitted first three lines
of Table 3) are plotted in Fig. \ref{flum}.

 The results described above
were checked in several ways. Here we exemplify the checks made, using
the i$^\prime$-band data, but equivalent results are obtained for the
g$^\prime$-band and r$^\prime$-band data.  The number of galaxies brighter
than i$^\prime=22$ above the line in Fig. \ref{cmd} which indicates the 
upper part of the red cluster sequence
is 72 for RX~J1552.2+2013 and $64 \pm 7$ for the mean of the two control (empty)
fields, indicating good agreement.  Note that, above this
line, the number of objects in the RX~J1552.2+2013 i$^\prime$ image  and in the empty
fields should match (apart, of course, from statistical fluctuations and
cosmic variance) given that these are expected to reflect the numbers
of background objects.  Moreover, if we plot a histogram of galaxy
counts for objects below the cluster red sequence line (see Fig. \ref{counts}),
the histogram flattens for i$^\prime \simgreat 20$. Since the field
counts increase continuously with magnitude, the number counts of group
members decreases.  Consequently, we are confident that the drop in the
luminosity function of the group RX~J1552.2+2013 is real.

Fossil groups have, by definition, a lack of bright galaxies because
of the selection function used to catalogue them. The bright-end of the
luminosity function of these systems is then known to be unusual, with
too few L* galaxies.  At the faint-end, nothing has been known, so far,
about the shape of the luminosity function of these groups. For RX~J1552.2+2013 we
just reach the magnitude of the dwarf upturn (the point where the curve
goes from being giant-dominated to dwarf-dominated), which in the Virgo
cluster is around m$_B$ = 17, corresponding
to M$_R$ = -18.5. Therefore, the steep downward slope at the faint-end
of the luminosity function shown in Fig. \ref{flum} is in part due to the lack
of dwarf galaxies in the sample. 
Such a pronounced dip in the luminosity
function has already been observed in several other systems' luminosity
functions and in particular in the composite luminosity function of
compact groups studied by \citet{Hunsberger}.
It is worth verifying whether this is a
peculiarity of RX~J1552.2+2013 or, otherwise, a common feature of fossil groups.

\subsection{Surface photometry of the brightest cluster galaxy}

\begin{figure}[!ht]
\includegraphics[width=\columnwidth]{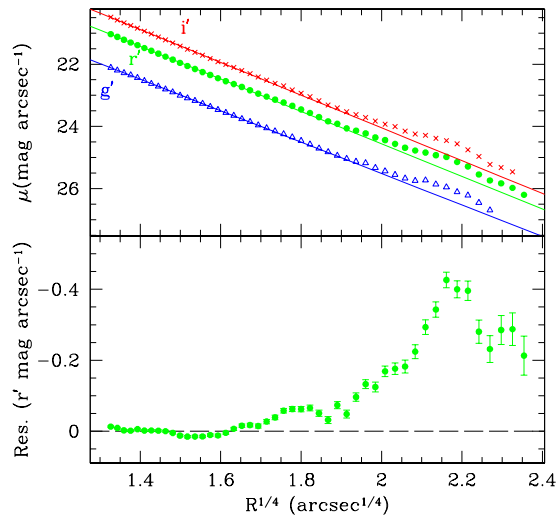}
\caption{\label{phot} {\it Upper panel} Photometric profile of the central
galaxy. We show the isophotal levels between a value of the semi-major axis of 
2.5\arcsec and that where the counts reach 1$\sigma$
of the background level (26.8, 26.3 and 25.6 mag arcsec$^{-2}$ for
g$^\prime$, r$^\prime$ and i$^\prime$ bands respectively) as a function of semi-major axis to
the power 1/4. The solid line is the best fit to the de Vaucoleurs profile
between 3.0 and 8.3\arcsec ($\mu_{r^{\prime}}=23.0$ mag arcsec$^{-2}$). {\it
Lower panel} residual between the actual r$^\prime$ band profile and the de Vaucoleurs
profile fit to the bright end.}
\end{figure}

In the upper panel of Fig. \ref{phot}, the azimutally averaged
photometric profile of the central galaxy of RX~J1552.2+2013 is shown.  
The surface photometry
was performed using the task {\sc ELLIPSE} in {\sc STSDAS/IRAF}, which
fits ellipses to extended object isophotes. We allowed the ellipticity
and position angle of the successive ellipses to change but the center
remained fixed.  The ellipse fitting was performed only in the deeper
r$^\prime$ image. For the other bands, the software measured the isophotal 
levels using the parameters estimated in the r$^\prime$ image.
There were a few small objects within the outer isophotes of the 
central galaxy which were masked during the profile fitting procedure.

We have fitted a r$^{1/4}$-law to the bright end of the galaxy profile, 
from well outside the seeing disk (3.0\arcsec) to a radius corresponding to
$\mu_{r^{\prime}}=23.0$ mag arcsec$^{-2}$ ($8.3\arcsec$).  
In the lower panel of
Fig. \ref{phot}, the residuals (data -- r$^{1/4}$-law model) for the r$^\prime$-band
data are shown. The light excess over the de-Vaucoleurs profile,  which is 
clearly detected in all three filters, is interpreted here
as due to an envelope. The
presence of this light envelope, in addition to the high luminosity of the
galaxy, strongly suggests that the central object of RX~J1552.2+2013 is a cD galaxy.

This result disagrees with a similar analysis done by \citet{jones03}
using a profile which reached similar isophotal levels.
On the other hand, we agree with \citet{jones03} that the subtraction
of the best-fitting elliptical model does not reveal multiple-nucleus,
shells or tidal tails.

\section{Discussion}

We list in Table \ref{allfossil} all fossil
groups studied to date.

\begin{deluxetable}{llllcl}
\tabletypesize{\scriptsize}
\tablewidth{0pt}
\tablecaption{Fossil group galaxies known to date \label{allfossil}}

\tablehead{\colhead{(1)} & \colhead{(2)} & \colhead{(3)} & \colhead{(4)} &
\colhead{(5)} & \colhead{(6)} \\
\colhead{Name} & \colhead{RA (2000)} & \colhead{DEC (2000)} &
\colhead{z} &
\colhead{L$_{X,bol} $ (10$^{42}$ h$_{50}$ $^{-2}$) ergs s$^{-1}$} & \colhead{Reference}}

\startdata
NGC 1132           & 02 52 51.8    & -01 16 29  & 0.0232   & ~~1.9   
& \citet{yoshioka04}   \\
RX~J0454.8-1806    & 04 54 52.2    & -18 06 56  & 0.0314   & ~~1.9   
& \citet{yoshioka04}   \\
ESO 306- G 017     & 05 40 06.7    & -40 50 11  & 0.035805 & 129     & \citet{sun04}        \\
RX~J1119.7+2126    & 11 19 43.7    & +21 26 50  & 0.061    & ~~1.7   & \citet{jones03}	  \\
RX~J1159.8+5531    & 11 59 51.4    & +55 32 01  & 0.0810   & ~22     & \citet{vikhlinin99}  \\
CL 1205+44         & 12 05 53.7    & +44 29 46  & 0.59     & 180     & \citet{ulmer05}	  \\
RX~J1256.0+2556    & 12 56 03.4    & +25 56 48  & 0.232    & ~61.    & \citet{jones03}	  \\
RX~J1331.5+1108    & 13 31 30.2    & +11 08 04  & 0.081    & ~~5.9   & \citet{jones03}	  \\
RX~J1340.6+4018    & 13 40 33.4    & +40 17 48  & 0.1710   & ~25     & \citet{vikhlinin99}  \\
RX~J1416.4+2315    & 14 16 26.9    & +23 15 32  & 0.137    & 220.    & \citet{jones03}      \\
RX~J1552.2+2013    & 15 52 12.5    & +20 13 32  & 0.136    & ~63     & \citet{jones03}	  \\
NGC 6034           & 16 03 32.1    & +17 11 55  & 0.0339   & ~~0.75  & \citet{yoshioka04}   \\
NGC 6482           & 17 51 48.8    & +23 04 19  & 0.013129 & ~~2.17  & \citet{khosroshahi04}\\
RX~J2114.3-6800    & 21 14 20.4    & -68 00 56  & 0.1300   & ~20     & \citet{vikhlinin99}  \\ 
RX~J2247.4+0337    & 22 47 29.1    & +03 37 13  & 0.199    & ~41     & \citet{vikhlinin99}  \\ 
\enddata
\small
\normalsize
\end{deluxetable}

     In Section 3.2 we showed that RX~J1552.2+2013 has a luminosity
function which has a dip at bright luminosities (by construction, since
fossil groups were chosen to be environments with no bright elliptical
galaxies besides the dominant elliptical) and, in addition, {\it the
luminosity function has a lack of faint galaxies  around M$_R$ = --18},
just before the dwarf upturn, as seen in other systems such as for the
Coma cluster.  The faint end of the luminosity function of RX~J1552.2+2013
is significantly different from that of other galaxy clusters of similar
mass. Indeed, for the 2dF and RASS--SDSS clusters, $\alpha \simeq 1.3$ in
the blue-band \citep{propris,popessoII}, instead of values in the range
--0.7 to --0.4 found here.  In a search in the literature,
we could find several examples of clusters which presented dips in
their luminosity functions, but they are mostly associated with dense,
dynamically well-evolved systems, such as  X-Ray emitting cD clusters
\citep[e.g.][]{Omar,valotto,mobasher}. On the other hand, a few
loose groups, like Leo I, has also been reported to have a dip in its
luminosity function at intermediate luminosities 
\citep[between --19.5 $< M_R <$ -- 16;][]{flint03},
and the luminosity function of compact groups of galaxies also 
show a similar dip (Hunsberger et al. 1996),
which suggests that there may be more than one mechanism in action,
for the depletion of these galaxies in different environments.

Fossil groups were suggested to be the end products of merging of L$^*$
galaxies in low-density environments \citep{jones03}.  The specific fossil
group studied here, however, does not constitute a low-density environment and,
in fact, is more similar to a galaxy cluster. The fairly high X-ray emission,
the large fraction of elliptical galaxies (most of the bright galaxies in
Fig. 1 are early-types), the radial velocity
distribution (Fig. 4), as well as the lack of obvious substructures, 
suggest that RX~J1552.2+2013 is probably virialized.

Under the merger scenario hypothesis, the mechanism usually claimed to
explain the lack of L$^*$ galaxies is dynamical friction. It works by
a deceleration of the orbital velocity of a galaxy,  where the galaxy loses its
kinetic energy to the pool of dark matter particles. This phenomenon causes the
galaxy to spiral towards the mass center and eventually to merge with the 
central galaxy. The frictional deceleration is proportional to the mass of 
the galaxy, so this process is expected to be 
more efficient for the
brightest cluster members. This process is probably more
efficient during the cluster collapse- when radial orbits may prevail- 
than after virialization, when the galaxy orbits are more isotropic
\citep[e.g.][]{Merritt85}.
Fossil groups/clusters appear to be an extreme case of this dynamical
friction scenario, since all
L$^*$ galaxies have merged and there was no 
further accretion of bright field galaxies into the system.
%This would
%require denser and ``colder'' initial conditions, actually more typical of
%compact groups \citep{Ishizawa} than of clusters.

Dynamical friction and subsequent merging are probably the processes
responsible for the lack of bright galaxies in the luminosity function of
fossil groups. However, it is interesting to note that these same physical
processes cannot be efficient enough for low-mass galaxies in order to 
explain the shortage of low-luminosity galaxies (around M$_R$ = --18 mag).

 An attractive explanation for the lack of faint galaxies in dense environments
is the dimming or even total disruption of these objects caused by a succession
of tidal encounters \citep{gnedin}. The debris of the disruption of such
galaxies is one of the possible sources of diffuse intra-cluster light 
and/or cD envelopes. Indeed, as shown in Section 3.3, the central galaxy
of RX~J1552.2+2013  has the light excess over its de Vaucouleurs profile that
characterizes cD galaxies. 

What is puzzling about fossil systems and RX~J1552.2+2013 in particular
is the high efficiency with which both mechanisms have acted. The
general case for clusters is to not have dips in the luminosity functions
\citep{propris,popessoII}. The exceptions are  the central regions of rich
clusters. For instance, the dip in the luminosity function of the Coma
cluster core is at R$\sim 17.0$ and B$\sim 18.0$ \citep{Trentham}, which
is close to the faintest limit of the luminosity functions presented here.

On the other hand, the disruption of  faint
galaxies is not instantaneous, requiring a few crossing times before tidal
heating can strip off less bound stars or even tear the galaxies 
apart. Therefore, the faint galaxies should be bound to these structures since
pre-virialization times. Moreover, the existence of a very compact and rich
core in this system should increase greatly the number of collisions, enhancing
the tidal stripping efficiency.

%In fact, since
%the efficiency of this mechanism increases with
%galaxy density, this is probably behind the relation between dwarf-to giant 
%ratio and local density \citep{ph98}, which indicates that dwarf galaxies 
%do tend to be less numerous in denser regions.

If this scenario is correct, we should not expect the same lack of faint
galaxies in less rich fossil groups, where the galaxy density and thus the
tidal encounters were much less frequent at early times. 
More observations are needed to test this scenario. 

One detail that should be kept in mind is that we have observed
just the inner part of RX~J1552.2+2013 ($\sim 1/3 R_{vir}$), region
where a few previous studies have found a lower dwarf-to-giant
ratio than that measured in the outskirts of the systems
\citep[e.g.][]{ph98}. \citet{popessoII}, however, using a very large
sample of 130 SDSS clusters with X-ray counterparts, have found only
a weak dependence of the dwarf-to-giant ratio with cluster-centric
distance. The fundamental parameter, in this case, seems to be the local
density \citep{driver98}, instead of the cluster--centric distance. And,
if we are to draw any conclusions based on the present knowledge of
fossil systems and the data we gathered, is that fossil systems had a
very dense origin.

   Considering the merging scenario, it is possible that the overluminous
galaxy has been formed within a substructure inside the already consolidated
larger structure.  In that case, one could think of a scenario where a compact
group was formed within a larger rich group, which would then have quickly
merged and would have left behind the brightest elliptical galaxy of what today
is seen as a fossil group. One weak argument against this scenario is that
compact groups  are not usually found within such massive structures, but
instead are more often surrounded by loose groups. This may, however, be a
selection effect in the compact group catalogues available in the literature,
which use an isolation criterion for group selection and may, therefore, select
against compact groups embedded in clusters.

  Another important point is that member galaxies in compact groups are
emerging as very old systems \citep{proctor04,claudia05}. This suggests two
possible scenarios for compact groups: they are either long-lived systems,
which have not had a major merger over a significant fraction of a Hubble time
\citep[scenario also supported by][]{zabludoff98} or they are systems which
merge so fast that they are not caught at intermediate stages of evolution. The
existence of a large population of fossil groups would support the latter
scenario of fast merging for compact groups, at least for  those within
clusters or rich groups.

One problem for the {\it merging origin} of fossil groups is their apparent much
higher mass-to-light ratios as compared to compact groups and
poor clusters. As explained earlier, recent work of \citet{yoshioka04} derive
very high values for the M/L of fossil groups and conclude that at least some
fossil groups are not end products of compact group evolution.  They suggest
that alternatively these objects are the massive tail of the elliptical galaxy
distribution, formed at high redshift ({\it fossil-elliptical origin}).  In the
present case, the system is definitely a cluster and not a single elliptical
galaxy.

The whole scenario of fossil group formation may become more clear
when more of these groups are studied spectroscopically.  Ongoing
determinations of the luminosity function for a large sample of fossil
groups and the detailed study of the properties of the brightest group
members, including the determination of their ages and metal abundances,
may also elucidate some of these
questions.  In addition, simulations such as those presented
recently by D'onghia et al. 2005 (astro-ph/0505544), but at higher resolution,
will allow a direct comparison of the observed luminosity function of
a fossil group and the expected luminosity function of merged groups.

%-------------------------------------------------------------------

\begin{acknowledgements}

We are grateful to Laurence Jones and Natalia Boris, for granting us access to
their data prior to publication.  We would like to thank the Gemini staff for
obtaining the observations, Elena D'Onghia, Daniele Pierini, 
Gast\~ao Lima Neto, Simon Driver and Roberto de
Propris for useful suggestions.  The authors would like to acknowledge support
from the Brazilian agencies FAPESP (projeto tem\'atico 01/07342-7), CNPq and CAPES. We
made use of the Hyperleda database and the NASA/IPAC Extragalactic Database
(NED). The latter is operated by the Jet Propulsion Laboratory, California
Institute of Technology, under contract with NASA.

\end{acknowledgements}

%-------------------------------------------------------------------

\bibliographystyle{aa}
\bibliography{abb,all,new}

%-------------------------------------------------------------------

\end{document}